\documentclass{article}
\usepackage{fullpage}
\usepackage{times}
\usepackage{amssymb}
\usepackage{graphicx}
\usepackage{amsmath}
\usepackage{amsthm}
\usepackage{subfigure}
\usepackage{bigints}
\usepackage{hyperref}

\theoremstyle{plain}
\theoremstyle{plain}
\theoremstyle{definition}

\begin{document}

\title {Varying-Alpha and K-Essence}
\author{Alexander A. H. Graham\footnote{Email: A.A.H.Graham@damtp.cam.ac.uk}\\
\emph{Department of Applied Mathematics and Theoretical Physics}\\
\emph{Centre for Mathematical Sciences}\\
\emph{University of Cambridge}\\
\emph{Wilberforce Road, Cambridge, CB3 0WA, UK}}
\date{\today}
\maketitle

\begin{abstract}
We introduce a model which allows the fine structure constant (alpha) to vary throughout space and time due to a coupling to a scalar field with a non-canonical kinetic structure. This provided a new extension of the Bekenstein-Sandvik-Barrow-Magueijo model of alpha variations. The background cosmology is studied in detail using dynamical systems techniques for a scalar field of ghost condensate type. We show generically that if the kinetic terms are chosen to allow an accelerated late-time attractor for the expansion scale factor then alpha will not asymptote to a constant at late times. 
\begin{description}
\item[PACS numbers:] 98.80.Es, 98.80.Bp, 98.80.Cq 
\end{description}
\end{abstract}

\maketitle


\section{Introduction}
Ever since Dirac developed what has become known as the large number coincidence \cite{dirac37, dirac38} physicists have been fascinated by the possibility that one or more of the constants of nature, and in particular the fine structure constant, \(\alpha\), may in fact be a slow-varying function of time (see Refs. \cite{bondi52, barrow81} for historical reviews). Today, the main motivations are two-fold. Theoretically, many theories of beyond the standard model physics predict that the observed constants of nature are in fact dynamical \cite{extra}. A good example of this is in higher dimensional models such as string theory: when one reduces the spacetime dimension down to four (in the manner of Kaluza-Klein) one will get several scalar fields coupling to gravity and all other matter fields. Indeed, in string theory there is a scalar field (the dilaton) coupling to the gauge fields even before dimensional reduction. Although the details are highly dependent on the compactification scheme one generically expects to find \(\mathcal{O}(100)\) scalar fields in the effective four-dimensional theory, many of which will directly couple to the Maxwell action. It would therefore seem prudent -- given that our understanding of such theories is insufficient to predict with confidence the precise low-energy, four-dimensional Lagrangian one expects -- to investigate phenomenological models which allow \(\alpha\) to vary in space and time.

An even more compelling reason is that there is some observational evidence to suggest that \(\alpha\) was lower in the distant past. The said observations are spectroscopic studies of quasars at redshifts \(z\sim\mathcal{O}(1)\). For over a decade the team of Webb et al have found evidence that \(\alpha\) was lower in the past by \(\Delta\alpha/\alpha=(\alpha (z)-\alpha _{0})/\alpha _{0}\sim-0.5\times 10^{-5}\),  where $\alpha _{0}$ is the value today, from careful studies of the fine-structure of atomic transitions seen in these quasars \cite{webb99, dzuba99, webb01, murphy01}. Despite much discussion in the literature these results have neither been refuted nor confirmed \cite{chand, murph}. The most recent measurements by this group found that for $z>1.8$ $\Delta {}\alpha /\alpha =(-0.74\pm 0.17)\times 10^{-5}$ using data for the Northern sky from the Keck telescope, but $\Delta{}\alpha /\alpha =(0.61\pm 0.20)\times 10^{-5}$ from data for the Southern sky from the VLT telescope (with some overlap), which would point to a large scale angular dipole of magnitude  $\sim0.6\times 10^{-5}$ \cite{webb11, king12}. Interestingly, the direction of the dipole is, in galactic coordinates \((l,b)\), \((l,b)\approx(330,-13)\), which is similar to the direction of other dipoles which have reported to have been observed in the CMB, in supernova data and in bulk flows \cite{mariano12, mariano12b}.

It is important to note that if one accepts the logical possibility that \(\alpha\) may have varied throughout cosmic history then there is no reason to suppose that the rate of \(\alpha\) variation is constant, indeed in most covariant models it is not. For this reason we cannot simply compare the measured variation in \(\alpha\) inferred from quasar measurements with the other constraints on \(\alpha\) variations obtained from atomic clock experiments today, from geological probes at \(z\sim0.1-0.4\) and from constraints based on the physics of the cosmic microwave background (CMB) and big bang nucleosynthesis (BBN) at \(z\approx {10^{3}}\) and \(z\approx {10^{8}}\) respectively (see Refs. \cite{uzan11, chiba11, murphy04} for detailed reviews). In fact, the relation between \(\Delta\alpha/\alpha\) at different epochs is always model dependent, and until a consistent theory of varying-\(\alpha\) is developed it is simply not possible to use constraints on \(\Delta\alpha/\alpha\) at one time to infer \(\Delta\alpha/\alpha\) at another time. It is also the case that in general one cannot use a local measurement of \(\Delta\alpha/\alpha\) (in, say, a stellar system) to infer the cosmological value of \(\Delta\alpha/\alpha\), since in any covariant theory of varying-\(\alpha\) the value of \(\alpha\) is sensitive to the gravitational potential. That said, one can prove for scalar field models of varying-\(\alpha\) under fairly general conditions \cite{shaw06, shaw06b} that the local value of \(\Delta\alpha/\alpha\) does indeed track the cosmological value of \(\Delta\alpha/\alpha\).

The first covariant model of varying-\(\alpha\) is due to Bekenstein \cite{bekenstein82}, who did so by coupling the Maxwell action to a massless scalar field. This was subsequently extended to a cosmological setting and studied in detail in Ref. \cite{sandvik02} by Sandvik, Barrow and Magueijo: we shall refer to it as BSBM theory. The action of this model is given by
\begin{equation} \label{1.1}
S=\int {d^{4}{x}}\sqrt{-g}\left( \frac{1}{2}R-\frac{1}{2}\omega \partial_{a}\phi {}\partial ^{a}\phi +e^{-2\phi}\mathcal{L}_{em}+\mathcal{L}_{m}\right),
\end{equation}
where $\omega $ is a coupling constant, $\mathcal{L}_{em}=-\frac{1}{4}F_{ab}F^{ab}$ is the usual electromagnetic Lagrangian and $\mathcal{L}_{m}$ denotes the Lagrangian for the other matter fields in the theory. In this model \(\alpha\) is given by
\begin{equation} \label{1.2}
\alpha =\alpha _{0}e^{2\phi },
\end{equation}
where $\alpha_{0}$ is a constant which may be taken as the present value of $\alpha $. Since $\mathcal{L}_{em}=\frac{1}{2}(E^{2}-B^{2})=0$ for radiation, variations in \(\alpha\) are driven solely by the electromagnetic energy of non-relativistic matter, parametrised by $\zeta _{m}=\mathcal{L}_{em}/\rho _{m}$ where $\rho _{m}$ is the energy density of non-relativistic matter. The cosmology of this theory has been studied by several authors and is now well understood, at least when $\zeta_{m}<0$ \cite{barrow02, barrow02a, barrow02b, barrow03}. In this case one finds that the scalar field has a negligible effect on the background expansion, and that the behaviour of \(\alpha\) is determined by the cosmological era one is in. One finds that in \(\Lambda\)CDM $\alpha $ does not grow in the radiation era, grows logarithmically with time in the dust era and asymptotes to a constant when the expansion starts to accelerate in the $\Lambda $-dominated era. The last behaviour, in particular, means that we would expect to find that the measured value of \(\Delta\alpha/\alpha\) today is extremely small even if it was observationally significant in the early universe. It is for this reason that the change in \(\alpha\) inferred from quasar observations is not in conflict with the terrestrial or the CMB and BBN constraints on \(\Delta\alpha/\alpha\). The model has been extended by the addition of a non-constant potential \cite{barrow08}, and by allowing the coupling to be a function of $\phi $ \cite{barrow12}. Recently, an extension of this model which allowed for both a potential and coupling function was studied, which also allowed the \(\zeta_{m}>0\) case to be investigated in detail via the methods of dynamical systems \cite{barrow13}. The perturbation theory of this generalised theory has also been investigated \cite{graham14}, and predicts the same growth of \(\alpha\) perturbations as in the original BSBM model. This model has also inspired similar models which allow the proton to electron mass ratio \(\mu\) to vary \cite{magueijo05}, and full electroweak models \cite{ew, ew2}. Recently a varying-\(\alpha\) model was proposed based on multiscale spacetimes \cite{calcagni14}. One of the conclusions of these studies was that it is difficult to construct a model which allows the universe to accelerate under the influence of a quintessence-like scalar field and simultaneously drives variations of alpha in a phenomenologically acceptable way. Generally, if we introduce a potential which drives acceleration then the late-time attractor for \(\alpha(t)\) will be some power-law (or faster) function of time \cite{barrow08}.

Of course, there is another logical possibility for driving acceleration. Instead of adding a potential we might modify the kinetic structure of the theory such that acceleration is driven by the kinetic energy of the scalar field. Such theories are well known in cosmology under the name of K-essence \cite{copeland06, debook}, since they are an alternative to standard quintessence theories in which the universe accelerates as the scalar field slowly rolls down its potential. They were first introduced into cosmology in Ref. \cite{mukhanov99} in the context of inflation, and soon studied \cite{chiba00, ap00, ap01} as an alternative to standard quintessence theories. Such theories can be well motivated from more fundamental theories like string theory \cite{debook}, and are able to provide interesting cosmologies where the late-time acceleration is driven by the kinetic, rather than potential, energy of a scalar field. The main attraction of these models is that they often have a better prospect for solving the coincidence problem than do standard quintessence theories, although to some extent this comes at the cost of potential problems involving causality and stability \cite{bonvin06}. Although there have been some studies of K-essence theories where the scalar is coupled to dust \cite{gumjudpai05}, and a study of a certain type of K-essence (the DBI model) which can couple naturally to the electromagnetic sector \cite{garousi05}, the general case with an electromagnetic coupling has not been investigated before. It is the aim of this paper to provide such a study. We shall examine the background dynamics of such a theory, and in particular study the special case where the kinetic term takes the form of the dilatonic ghost condensate model. We will find that in general the late-time attractor is a scalar-dominated state with \(\alpha\) growing at least as fast as a power-law of time. These models are thereby observationally problematic in the same way as quintessence models of varying-\(\alpha\) are.

The outline of this paper is as follows. The model is introduced in section 2 and its basic details are discussed before we specialise to the cosmology of the model in an FRW spacetime in section 3. Sections 4 and 5 then study the late-time cosmological dynamics in detail when the scalar field is of ghost condensate and dilatonic ghost condensate type. We show that in both cases \(\alpha\) never asymptotes to a constant, and argue in section 6 that this is a generic feature of these models. We conclude in section 7.

From now on we will employ units in which \(c=\hbar=8\pi{G}=1\) unless otherwise stated.

\section{K-essence models of varying $\alpha$}
The model we study in this paper is defined by the action
\begin{equation} \label{2.1}
S=\int{}d^{4}x\sqrt{-g}\left( \frac{1}{2}R+P(\phi,X)+e^{-2\phi/\sqrt{\omega}}\mathcal{L}_{em}+\mathcal{L}_{m}\right),
\end{equation}  
where \(X=-\frac{1}{2}\nabla_{a}\phi\nabla^{a}\phi\) and \(\omega\) is a constant which plays the same role as the constant \(\omega\) in the original BSBM model. In particular, the coupling between the scalar field and the electromagnetic sector vanishes in the limit \(\omega\rightarrow \infty\). Note that there is no loss of generality in restricting to an exponential coupling, since the general case can always be reduced to the action \eqref{2.1} by a field redefinition. The theory of Ref. \cite{barrow13} is recovered by the choice \(P=\omega(\phi)X-V(\phi)\). In this model the fine structure constant is given by
\begin{equation} \label{2.2}
\alpha=\alpha_{0}e^{2\phi/\sqrt{\omega}}. 
\end{equation}

By varying the action with respect to the metric it is easy to see that the Einstein equations are
\begin{equation} \label{2.3}
G_{ab}=T_{ab}^{m}+T_{ab}^{\phi }+e^{-2\phi/\sqrt{\omega}}T_{ab}^{em},  
\end{equation}
where \(T_{ab}^{em}\) and \(T_{ab}^{m}\) are the usual energy-momentum tensors of the electromagnetic and any uncoupled matter fields respectively, and the energy-momentum tensor of the scalar field, \(T_{ab}^{\phi}\), is  
\begin{equation} \label{2.4}
T_{ab}^{\phi}=P_{,X}\partial_{a}\phi\partial_{b}\phi+Pg_{ab}. 
\end{equation}
Varying the action with respect to \(\phi\) gives the scalar field equation of motion: 
\begin{equation} \label{2.5}
P_{,X}\Box\phi+\nabla^{a}\phi(\nabla_{a}\phi{}P_{,X\phi}-\nabla^{b}\phi\nabla_{a}\nabla_{b}\phi{}P_{,XX})+P_{,\phi}=\frac{2}{\sqrt{\omega}}e^{-2\phi/\sqrt{\omega}}\mathcal{L}_{em}. 
\end{equation}
The Maxwell equations are essentially unchanged from those in Ref. \cite{barrow13}, and in any event they will not be needed here. We will usually rewrite the right hand side of Eq. \eqref{2.5} by defining a dimensionless parameter $\zeta $ by
\begin{equation} \label{2.55}
\zeta =\frac{\mathcal{L}_{em}}{\rho },
\end{equation}
and $\zeta _{m}=\mathcal{L}_{em}/\rho _{m}$ for its cosmological value, where $\rho _{m}$ is the energy density of non-relativistic matter. This is convenient because non-relativistic matter is the only source in Eq. \eqref{2.5}. Clearly, \(\zeta\) may take values in the range \(-1\leq {}\zeta {}\leq 1\). Although this parameter will in general vary from material to material the cosmological value should be approximately constant, at least over the timescales we consider. In this paper we always make this assumption. We shall also allow for either value of \(\zeta _{m}\) in this paper -- see Refs. \cite{sandvik02, barrow13, bekenstein02} for a discussion on its likely value and sign.

Introducing non-canonical kinetic terms into a theory is not a step one should entertain lightly, and it potentially comes with various serious risks for the theories viability. In particular, the theory may become ghost-like (admit states of negative energy) and the theory may allow for superluminal propagation, which can entail a loss of causality. Fortunately this is a well studied problem for this type of theory, and it is known how to avoid such issues \cite{mukhanov99b, hamed04, piazza04, lim05}. Classical stability of the theory is guaranteed if the sound speed of the theory,
\begin{equation} \label{2.6}
c_{s}^{2}=\frac{P_{,X}}{P_{,X}+2XP_{,XX}}, 
\end{equation}
is always positive. This is also the same condition for the theory to admit a well-posed initial value formulation \cite{lim05, rendall06}, which again is essential for the theory to make sense classically. Quantum stability, which requires that the perturbed Hamiltonian about a background solution is positive, demands that the following conditions be meet: 
\begin{equation} \label{2.7}
P_{,X}\geq0,\  \mbox{ and }\ P_{,X}+2XP_{,XX}\geq0. 
\end{equation}
Quantum stability implies classical stability, but the converse is false. Causality is manifestly maintained if \(c_{s}^{2}\leq1\), although it has been argued \cite{babichev08} that these models may still be causally consistent even when the sound speed is superluminal (see Ref. \cite{ellis07} for a contrary opinion). Provided that the stability conditions \eqref{2.7} are satisfied then \(c_{s}^{2}\leq1\) iff \(XP_{,XX}\geq0\).

As is well known, any covariant model of varying \(\alpha\) must necessarily violate the weak equivalence principle to some degree, and this theory is no exception \cite{dicke62, bekenstein82}. Usually, one parameterises violations of the weak equivalence principle in terms of the E\"{o}tv\"{o}s parameter $\eta $: present experimental limits are of the order \(\eta\leq\mathcal{O}(10^{-13})\) \cite{wepviolation08}. One can calculate the  E\"{o}tv\"{o}s parameter by computing the Newtonian limit of the theory, which can be done using similar steps to the calculation in Ref. \cite{barrow13}. One finds that the scalar field in this limit obeys the same equation as in Ref. \cite{barrow13}. This means predictions for the E\"{o}tv\"{o}s parameter are the same as in BSBM theory, and so this model violates the weak equivalence principle to the same extent as the BSBM model does \cite{bekenstein82, bekenstein02, maj, damour03}. 

It is worth noting that there is a potential loop-hole in these results if the theory admits what is known as the Vainshtein mechanism \cite{vainshtein72}. This is when non-linearities in the theory dominate around compact objects due to an unusual kinetic structure of the theory, which causes a suppression of fifth force effects and consequently reduces the violation of the weak equivalence principle one expects. It has been shown \cite{brax13} that K-essence theories can potentially admit Vainshtein screening, although the results in that paper are not directly relevant here, since they assumed the scalar field couples to the trace of the energy-momentum tensor.

\section{Cosmology of the model}
We now study the behaviour of this model in a Friedmann-Robertson-Walker (FRW) universe. This means the spacetime metric will take the form
\begin{equation} \label{3.1}
ds^{2}=-dt^{2}+a^{2}(t)\left[ \frac{dr^{2}}{1-kr^{2}}+r^{2}(d\theta^{2}+\sin ^{2}\theta {}d\phi ^{2})\right],
\end{equation}
where \(a(t)\) is the scale factor and \(k\) the spatial curvature. Note that in this paper we choose to normalise the scale factor so that today it is normalised to unity: \(a(t_{0})=1\), where \(t_{0}\) is the current age of the universe. Given this metric it is straightforward to show that the scalar field equation of motion and the Einstein equations in an FRW spacetime take the form
\begin{align}
&\ddot{\phi}(P_{,X}+\dot{\phi}^{2}P_{,XX})+P_{,X\phi}\dot{\phi}^{2}+3HP_{,X}\dot{\phi}-P_{,\phi}=-\frac{2}{\sqrt{\omega}}e^{-2\phi/\sqrt{\omega}}\zeta_{m}\rho_{m}, \label{3.2}
\\
&H^{2}=\frac{1}{3}\left( \rho _{m}(1+|\zeta_{m}|e^{-2\phi/\sqrt{\omega}})+\rho _{r}e^{-2\phi/\sqrt{\omega}}+(\dot{\phi}^{2}P_{,X}-P)\right)-\frac{k}{a^{2}}, \label{3.3}
\\
&\dot{H}=-\frac{1}{2}\rho _{m}(1+|\zeta _{m}|e^{-2\phi/\sqrt{\omega}})-\frac{2}{3}\rho _{r}e^{-2\phi/\sqrt{\omega}}-\frac{1}{2}\dot{\phi}^{2}P_{,X}+\frac{k}{a^{2}}. \label{3.4}
\end{align}
These are supplemented by the usual continuity equation for dust \(\rho_{m}\), while the radiation \(\rho_{r}\) obeys the modified continuity equation
\begin{equation}\label{3.41}
\dot{\rho}_{r}+4H\rho_{r}=\frac{2\dot{\phi}\rho_{r}}{\sqrt{\omega}}, 
\end{equation}
due to the explicit coupling between relativistic matter and the scalar field. These equations are not easy to solve except via highly restrictive assumptions. To study their dynamics we shall formulate these equations as a dynamical system \cite{glendinning94}. This is a well known method which has been applied widely in cosmology \cite{copeland06, debook, wainwright97, barrow86, goliath99}. The approach we take is based largely on Refs. \cite{piazza04, copeland98}. Although Eqs. \eqref{3.2}-\eqref{3.4} are valid for an arbitrary \(P(\phi,X)\) unfortunately it is not easy to cast these equations into an autonomous form for a general \(P(\phi,X)\). Instead, we will restrict to the choice
\begin{equation} \label{3.5}
P=-X+\frac{e^{\lambda\phi}X^{2}}{M^{4}}-V(\phi), 
\end{equation}
where \(\lambda\) and \(M\) are constants. This is the dilatonic ghost condensate model \cite{piazza04}, which is a fairly straightforward extension of the well known ghost condensate proposal \cite{hamed04} (we have allowed a potential term for greater generality). It reduces to the ghost condensate model with a potential when \(\lambda=0\). The idea of this model is that while the first term in the action is ghost-like it can be stabilised by the presence of the higher derivative second term while still admitting self-accelerating solutions. This can be seen explicitly by evaluating the stability constraints \eqref{2.7} for this model: stability is guaranteed if
\begin{equation} \label{3.51}
\frac{e^{\lambda\phi}X}{M^{4}}\geq\frac{1}{2}.  
\end{equation}
One can also check that causality follows automatically as the sound speed \(c_{s}^{2}\leq1\) (note that in an FRW spacetime \(X\geq0\)). One might say that this theory is stable provided it always has a minimum amount of kinetic energy. In any event, whatever its theoretical attractions, or otherwise, it is a good toy model of cosmologies which accelerate under the influence of the kinetic energy of the scalar field rather than through a potential. For this choice of action Eqs. \eqref{3.2}-\eqref{3.4} become
\begin{align}
&\ddot{\phi}\left(\frac{3e^{\lambda\phi}\dot{\phi}^{2}}{M^{4}}-1\right)+3H\dot{\phi}\left(\frac{e^{\lambda\phi}\dot{\phi}^{2}}{M^{4}}-1\right)+\frac{3\lambda{}e^{\lambda\phi}\dot{\phi}^{4}}{4M^{4}}+V'=-\frac{2}{\sqrt{\omega}}e^{-2\phi/\sqrt{\omega}}\zeta_{m}\rho_{m} \label{3.6}
\\
&H^{2}=\frac{1}{3}\rho _{m}(1+|\zeta_{m}|e^{-2\phi/\sqrt{\omega}})+\frac{1}{3}\rho _{r}e^{-2\phi/\sqrt{\omega}}-\frac{1}{6}\dot{\phi}^{2}+\frac{1}{4}\frac{e^{\lambda\phi}\dot{\phi}^{4}}{M^{4}}+\frac{V}{3}-\frac{k}{a^{2}}, \label{3.7}
\\
&\dot{H}=-\frac{1}{2}\rho _{m}(1+|\zeta _{m}|e^{-2\phi/\sqrt{\omega}})-\frac{2}{3}\rho _{r}e^{-2\phi/\sqrt{\omega}}+\frac{1}{2}\dot{\phi}^{2}-\frac{e^{\lambda\phi}\dot{\phi}^{4}}{2M^{4}}+\frac{k}{a^{2}}. \label{3.8}
\end{align}
From now on we will simplify the analysis by restricting to spatially flat cosmologies -- the case with non-zero curvature can also be done without much additional difficulty \cite{goliath99, barrow13}. To cast them into autonomous form we follow the strategy of Ref. \cite{debook}. We first define the autonomous variables
\begin{equation} 
x_{1}=\frac{\dot{\phi}}{\sqrt{6}H},\ x_{2}=\frac{\sqrt{V}}{\sqrt{3}H},\ x_{3}=\frac{\sqrt{\rho_{m}|\zeta_{m}|}e^{-\phi/\sqrt{\omega}}}{\sqrt{3}H},\ x_{4}=\frac{\dot{\phi}e^{\lambda\phi/2}}{\sqrt{2}M^{2}},\ x_{5}=\frac{\sqrt{\rho_{r}}e^{-\phi/\sqrt{\omega}}}{\sqrt{3}H}. \label{3.9}
\end{equation}
We also define \(x_{0}=\sqrt{\rho_{m}}/\sqrt{3}H\), which is gotten from the constraint equation
\begin{equation} \label{3.10}
1=x_{0}^{2}+x_{3}^{2}+x_{2}^{2}+x_{1}^{2}(3x_{4}^{2}-1)+x_{5}^{2}. 
\end{equation}
Notice that \(x_{0}\), \(x_{2}\), \(x_{3}\), \(x_{5}\) must always be positive, while the sign of \(x_{1}\) and \(x_{4}\) can in principle take either sign. The stability conditions \eqref{2.7} are satisfied provided that \(|x_{4}|\geq\frac{1}{\sqrt{2}}\), which implies that \(|x_{1}|\leq\sqrt{2}\). In these variables \(\alpha\) is given by
\begin{equation} \label{3.11}
\frac{\alpha}{\alpha_{0}}=|\zeta_{m}|\left(\frac{x_{0}}{x_{3}}\right)^{2}, 
\end{equation}
while the Hubble parameter \(H\) is given by
\begin{equation} \label{3.12}
\left(\frac{H}{H_{0}}\right)^{2}=\left(\frac{x_{0,0}}{x_{0}}\right)^{2}e^{-3N}, 
\end{equation}
where \(H_{0}\) and \(x_{0,0}\) is the Hubble parameter and the value of \(x_{0}\) today. We have defined a logarithmic time coordinate \(N=\ln{}a(t)\). This is the time coordinate we use in the dynamical systems analysis, as it leads to simpler evolution equations for the autonomous variables. Notice that for ever expanding cosmologies with a big bang the interval \(t\in[0,\infty)\) is mapped to \(N\in(-\infty,\infty)\). If the potential is non-constant then we also need to define 
\begin{equation} \label{3.13}
\lambda_{V}=-\frac{{V}'}{{V}},\ {}\ \Gamma_{V}=\frac{{V}{V}''}{{V}'^{2}}. 
\end{equation}
Using Eqs. \eqref{3.6}-\eqref{3.8} the evolution equations for these variables become
\begin{align}
&\frac{dx_{1}}{dN}=\frac{1}{2}x_{1}(3+3x_{1}^{2}(x_{4}^{2}-1)-3x_{2}^{2}+x_{5}^{2})-\frac{1}{6x_{4}^{2}-1}\left(3x_{1}(2x_{4}^{2}-1)+3\sqrt{\frac{3}{2}}\lambda{}x_{4}^{2}x_{1}^{2}-\sqrt{\frac{3}{2}}\lambda_{V}x_{2}^{2}+\sqrt{\frac{6}{\omega}}x_{3}^{2}\hat{\zeta}_{m}\right), \label{3.14}
\\
&\frac{dx_{2}}{dN}=-\sqrt{\frac{3}{2}}\lambda_{V}x_{1}x_{2}+\frac{1}{2}x_{2}(3+3x_{1}^{2}(x_{4}^{2}-1)-3x_{2}^{2}+x_{5}^{2}), \label{3.15}
\\
&\frac{dx_{3}}{dN}=-\sqrt{\frac{6}{\omega}}x_{1}x_{3}+\frac{1}{2}x_{3}(3x_{1}^{2}(x_{4}^{2}-1)-3x_{2}^{2}+x_{5}^{2}), \label{3.16}
\\
&\frac{dx_{4}}{dN}=\sqrt{\frac{3}{2}}\lambda{}x_{1}x_{4}-\frac{x_{4}}{6x_{4}^{2}-1}\left(3(2x_{4}^{2}-1)+3\sqrt{\frac{3}{2}}\lambda{}x_{4}^{2}x_{1}-\sqrt{\frac{3}{2}}\lambda_{V}\frac{x_{2}^{2}}{x_{1}}+\sqrt{\frac{6}{\omega}}\hat{\zeta}_{m}\frac{x_{3}^{2}}{x_{1}}\right), \label{3.17}
\\
&\frac{dx_{5}}{dN}=\frac{1}{2}x_{5}(-1+3x_{1}^{2}(x_{4}^{2}-1)-3x_{2}^{2}+x_{5}^{2}), \label{3.18}
\end{align}
where we write \(\hat{\zeta}_{m}=\zeta_{m}/|\zeta_{m}|\) for the sign of \(\zeta_{m}\). These are supplemented by the equation
\begin{equation} \label{3.19}
\frac{d\lambda_{V}}{dN}=-\sqrt{6}\lambda_{V}^{2}(\Gamma_{V}-1)x_{1}
\end{equation}
if the potential is neither a constant or an exponential. For convenience it is also useful to define the parameter \(x_{\phi}\) by
\begin{equation} \label{3.20}
x_{\phi}=x_{1}^{2}(3x_{4}^{2}-1). 
\end{equation}
It is the total contribution of the scalar field to the density parameter. Note that the subspaces \(x_{2}=0\), \(x_{3}=0\), \(x_{4}=0\), \(x_{5}=0\) and \(\lambda_{V}=0\) are all invariant submanifolds of the solution space, so we are always free to study the system with these restrictions. A full phase-plane analysis of Eqs. \eqref{3.14}-\eqref{3.18} could in principle be done, but is somewhat tedious due to the large number of variables. Instead, since we are mainly interested in the late-time behaviour we will ignore radiation and focus on a universe containing only dust and a scalar field (so \(x_{5}=0\)). Since we are primarily interested in the case where acceleration is driven by the scalar fields kinetic energy rather than the potential we will also set \(V=0\) (\(x_{2}=0\)).

\section{Ghost condensate model}
We first study the case when the scalar field is of ghost condensate type, that is we set \(\lambda=0\). This case is of course somewhat easier than the analysis of the general dilatonic ghost condensate model, and is probably of more interest in any event. Table 1 lists the physical stationary points for this system. It should be noted these are not the only stationary points. In addition to these there is also another dust-dominated stationary point with \((x_{1},x_{3},x_{4})=(0,0,0)\) and a point with all of the variables non-zero:
\begin{equation} \label{4.1}
(x_{1},x_{3},x_{4})=\left(-\frac{1}{2}\sqrt{\frac{3\omega}{2}},\sqrt{\frac{2}{\hat{\zeta}_{m}}\left(\frac{3\omega}{8}-2\right)},\sqrt{1-\frac{8}{3\omega}}\right).  
\end{equation}
The first violates the stability bound, while it can be shown the second cannot satisfy the stability bound with \(x_{0}^{2}\geq0\).  
\begin{table} [t]
\begin{center}
  \begin{tabular}{| l || c | c | c | c | c | c | p{3.5cm} | }
    \hline
   Stationary point (SP) & \(x_{0}\) & \(x_{1}\) & \(x_{3}\) & \(x_{4}\) & \(x_{\phi}\) & Existence & Stability  \\ \hline
   \(1\) & \(1\) & \(0\) & \(0\) & \(\frac{1}{\sqrt{2}}\) & \(0\) & All \(\omega\), \(\hat{\zeta}_{m}\) & Saddle point \\ 
    \(2\) & \(0\) & \(\sqrt{2}\) & \(0\) & \(\frac{1}{\sqrt{2}}\) & \(1\) & All \(\omega\), \(\hat{\zeta}_{m}\) & Stable node \\  
    \(3\) & \(0\) & \(-\sqrt{2}\) & \(0\) & \(\frac{1}{\sqrt{2}}\) & \(1\) & All \(\omega\), \(\hat{\zeta}_{m}\) & Stable node \(\omega>\frac{16}{3}\), saddle point \(\omega<\frac{16}{3}\) \\  \hline

\end{tabular} 
\end{center}
\caption{Stationary points for a universe with dust and K-essence scalar field of ghost condensate type, with zero potential (excluding points violating the stability bound). The variables are defined in Eqs. \eqref{3.9} and \eqref{3.20}. Note that the sign of \(x_{4}\) is not fixed from the equations -- we have taken the positive root. Point 2 is the physically relevant attractor of this system.}
\end{table}
The first physical point is the Einstein-de Sitter universe. Since the eigenvalues for this point are \((\frac{3}{2},0,-3)\) it is a saddle point at late times. The second point is a scalar-dominated state with dynamics
\begin{equation} \label{4.2}
\phi(t)=\phi_{0}+M^{2}t, \ {}\ a(t)=a_{0}e^{\frac{M^{2}t}{2\sqrt{3}}}. 
\end{equation}
The eigenvalues of this point can be easily shown to be \((-3,-3,-\frac{3}{2}-2\sqrt{\frac{3}{\omega}})\), and so it is always the attractor for this system. This is the same attractor as in the ghost condensate model with zero coupling to the electromagnetic sector \cite{hamed04}, and so it is clear that the coupling of the ghost condensate field to the electromagnetic sector has no effect on the end-state of the evolution. In fact, the only effect is that this scalar-dominated attractor is approached even faster compared with the uncoupled case. This is also confirmed through numerical simulations. Figure 1 shows numerical simulations of this system for different signs of the coupling \(\zeta_{m}\). In both cases the scalar-dominated attractor is reached quickly.

The last point is another scalar-dominated end state with similar behaviour to the third stationary point, except that \(\phi\) is a decreasing function of time:
\begin{equation} \label{4.21}
\phi(t)=\phi_{0}-M^{2}t, \ {}\ a(t)=a_{0}e^{\frac{M^{2}t}{2\sqrt{3}}}. 
\end{equation}
Its eigenvalues are \((-3,-3,-\frac{3}{2}+2\sqrt{\frac{3}{\omega}})\), so it is only an attractor when \(\omega>\frac{16}{3}\).

Although the coupling does not affect the background expansion rate, it does though mean that at late times \(\alpha\) evolves as an exponential of time:
\begin{equation} \label{4.3}
\alpha(t)=\alpha_{0}e^{\frac{2M^{2}t}{\sqrt{\omega}}}. 
\end{equation}
Such a fast variation of \(\alpha\) is clearly phenomenologically undesirable, and rules against this model as a viable cosmology. 

\begin{figure}[htp]
  \begin{center}
    \subfigure[\(\hat{\zeta}_{m}=1\)]{\label{fig:edge-a}\includegraphics[scale=1, width=6.5cm, height=3.5cm]{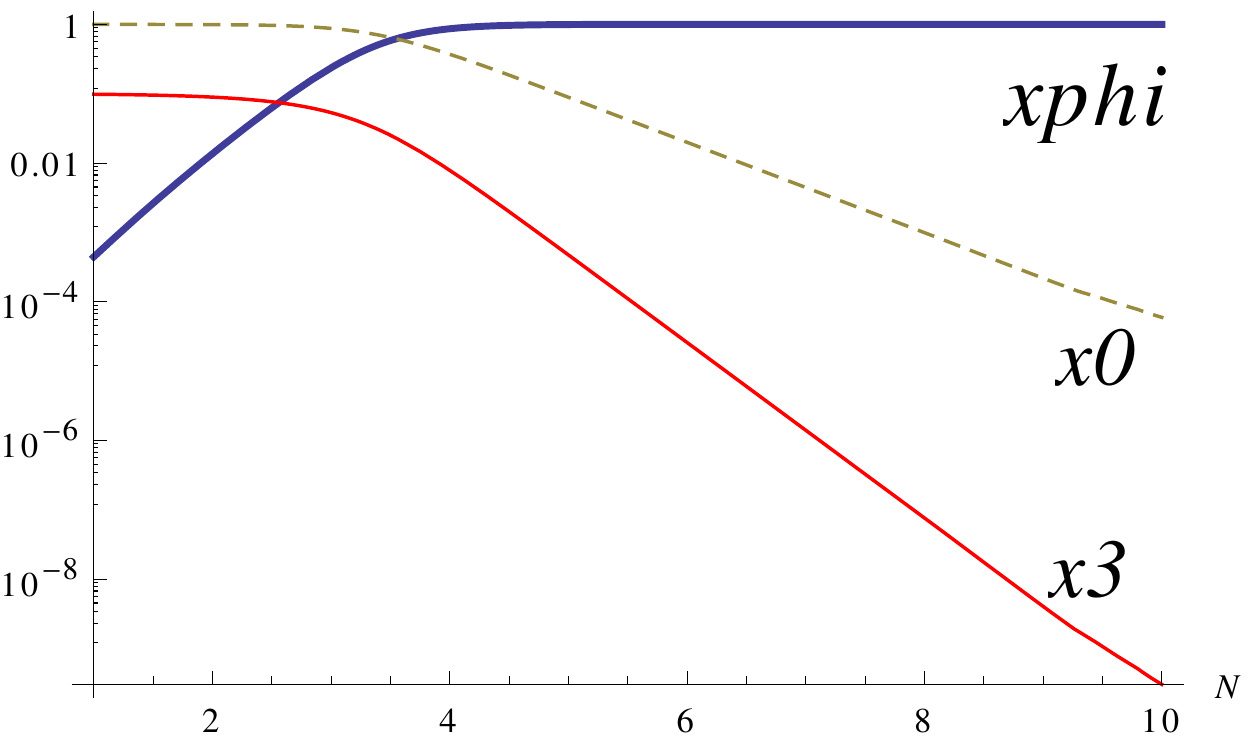}}
    \subfigure[\(\hat{\zeta}_{m}=-1\)]{\label{fig:edge-b}\includegraphics[scale=1, width=6.5cm, height=3.5cm]{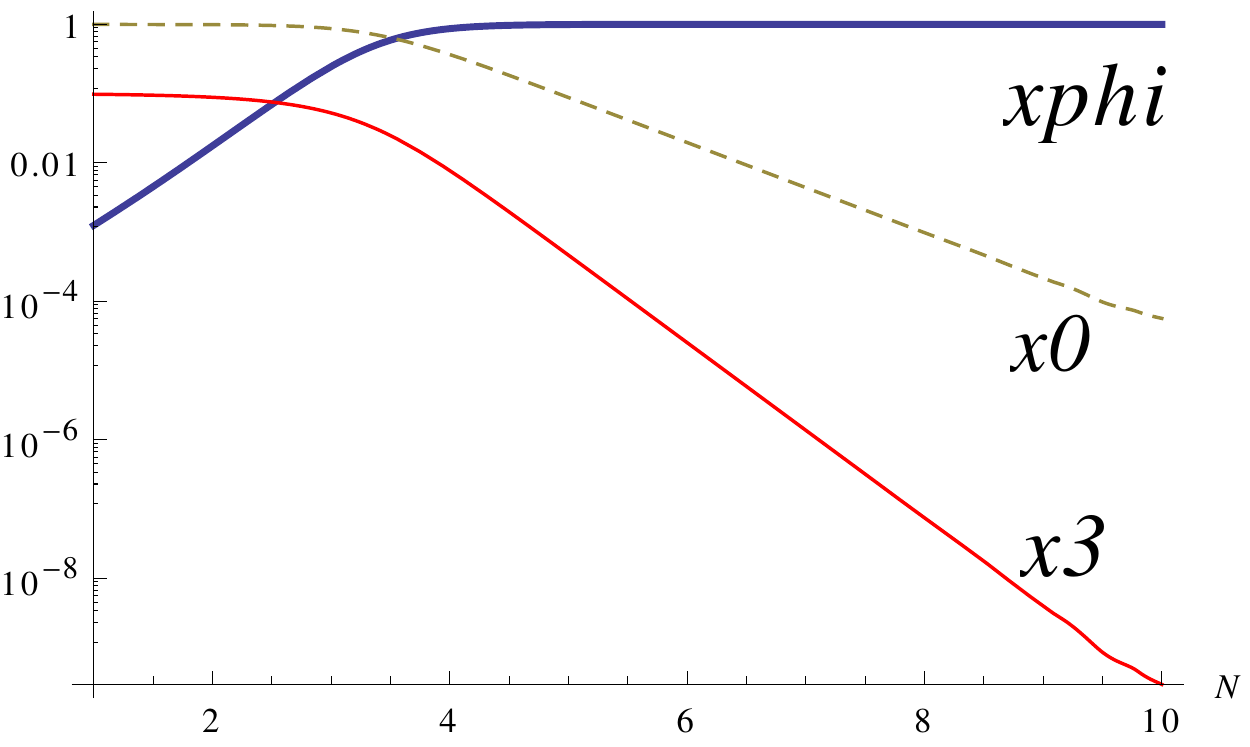}}
  \end{center}
  \caption{Simulations for a universe with dust and a ghost condensate scalar field, with \(\omega=6\) and \(\lambda=0.0\) in both cases. The thick blue curve is the magnitude of \(x_{\phi}=x_{1}^{2}(3x_{4}^{2}-1)\), the red of \(x_{3}\) and the dashed yellow of \(x_{0}\). For both simulations we choose initial conditions \(x_{1}(0)=0.01\), \(x_{3}(0)=0.1\) and \(x_{4}(0)=0.8\). In both cases the solution quickly becomes dominated by the kinetic energy of the scalar field and approaches an uncoupled, K-essence attractor solution. The sign of \(\zeta_{m}\) clearly has little effect on the dynamics.}
  \label{fig:edge}
\end{figure}

\section{Dilatonic ghost condensate model}
We now study the cosmology of the general dilatonic ghost condensate model. Table 2 shows the physical stationary points for this system. As with the ghost condensate model there are several other stationary points which violate the stability bound. This includes analogues of points 1 and 2 in table 2 with \(x_{4}^{2}<\frac{1}{2}\), and a generalisation of the stationary point in Eq. \eqref{4.1} which again can be shown to violate the stability bound for physical values of the parameters.
\begin{table} [t]
\begin{center}
  \begin{tabular}{| l || c | c | c | c | c | c | p{3.5cm} | }
    \hline
    SP & \(x_{0}\) & \(x_{1}\) & \(x_{3}\) & \(x_{4}\) & \(x_{\phi}\) & Existence & Stability  \\ \hline
    \(1\) & \(1\) & \(0\) & \(0\) & \(\frac{1}{\sqrt{2}}\) & \(0\) & All \(\omega\), \(\hat{\zeta}_{m}\) & Saddle point \\ 
    \(2\) & \(0\) & \(-\frac{\sqrt{6}}{4}\lambda\left(1-\sqrt{1+\frac{16}{3\lambda^{2}}}\right)\) & \(0\) & \(\sqrt{\frac{1}{2}+\frac{\lambda^{2}}{16}\left(1+\sqrt{1+\frac{16}{3\lambda^{2}}}\right)}\) & \(1\) & All \(\omega\), \(\hat{\zeta}_{m}\) & Stable node \(0\leq\lambda<\sqrt{3}\) \\  
    \(3\) & \(\sqrt{1-\frac{3}{\lambda^{2}}}\) & \(\sqrt{\frac{3}{2}}\frac{1}{\lambda}\) & \(0\) & \(1\) & \(3/\lambda^{2}\) & \(\lambda^{2}>3\) & Stable node \(\lambda>\sqrt{3}\) \\ \hline
   
\end{tabular} 
\end{center}
\caption{Stationary points for a universe with dust and K-essence scalar field of dilatonic ghost condensate type, with zero potential (excluding points violating the stability bound). The variables are defined in Eqs. \eqref{3.9} and \eqref{3.20}. Note that the sign of \(x_{4}\) is not fixed from the equations -- we have taken the positive root. Point 2 is the attractor when \(0\leq\lambda<\sqrt{3}\), while when \(\lambda>\sqrt{3}\) point 3 is the attractor.}
\end{table}
The first physical point is the Einstein-de Sitter universe. It has the same eigenvalues as in the ghost condensate case, so is a saddle point at late times. The second point is a scalar-dominated solution with solution
\begin{equation} \label{5.1}
\phi(t)=\phi_{0}+\frac{2}{\lambda}\ln\left(\frac{\beta(\lambda){}M^{2}\lambda{}t}{\sqrt{2}}\right), \ {}\ a(t)=a_{0}{t}^{\frac{\sqrt{2}}{\sqrt{3}\lambda{}\alpha(\lambda)}}, 
\end{equation} 
where \(\alpha(\lambda)\) and \(\beta(\lambda)\) are the values of \(x_{1}\) and \(x_{4}\) at the value of the stationary points respectively. It exists for all values of \(\lambda\), though it is only an accelerated state when \(\lambda^{2}<2/3\). Although the eigenvalues for this point are somewhat complex one can show that it is an attractor when \(0\leq\lambda<\sqrt{3}\) for all values of \(\omega\), while when \(\lambda>\sqrt{3}\) it is a saddle point. When \(\lambda<0\) there is a small range, depending upon the value of \(\omega\), for which this point is stable; it is never an attractor when \(\lambda<-\sqrt{3}\). In this case the system evolves to a similar, unstable attractor point which has \(x_{4}^{2}<\frac{1}{2}\).

The final point is a scaling solution in which the ratio of the density parameter of the dust to the scalar field is fixed:
\begin{equation} \label{5.2}
\frac{x_{0}^{2}}{x_{\phi}}=\frac{\lambda^{2}}{3}-1. 
\end{equation}
This point only exists when \(\lambda^{2}>{3}\). The scale factor is of the form of an Einstein-de Sitter universe, but with the scalar field growing logarithmically in time:
\begin{equation} \label{5.3}
\phi(t)=\phi_{0}+\frac{2}{\lambda}\ln\left(\frac{M^{2}\lambda{}t}{\sqrt{2}}\right), \ {}\ a(t)=a_{0}t^{\frac{2}{3}}. 
\end{equation}
The eigenvalues of this point can be shown to be
\begin{equation} \label{5.4}
\frac{3}{20\lambda}(-5\lambda{}\pm\sqrt{15(8-\lambda)}),\ {}\ \frac{-3}{\sqrt{\omega}}\frac{1}{\lambda}, 
\end{equation}
which implies this is an attractor only when \(\lambda>\sqrt{3}\) (when \(\lambda<-\sqrt{3}\) it is clearly a saddle point). We see that when \(\lambda>0\) the system evolves either to a purely scalar-dominated solution or a scaling solution. In both of these cases \(\phi\) does not asymptote to a constant but grows logarithmically with time. This means of course that in either case \(\alpha\) must necessarily grows as a power law of time:
\begin{equation} \label{5.5}
\alpha(t)=\alpha_{0}t^{\frac{4}{\lambda\sqrt{\omega}}}. 
\end{equation}
This behaviour can again be seen easily from numerical simulations. Figure 2 shows numerical simulations of this system for the value \(\lambda=\frac{1}{2}\) and for different signs of the coupling \(\zeta_{m}\). In both cases the scalar-dominated attractor is reached quickly, although somewhat slower than in the ghost condensate simulation of figure 1 as we would expect. Figure 3 also shows numerical simulations of this system when \(\lambda=2\), and clearly shows the development of the scaling-solution attractors which are again independent of the sign of the coupling.

\begin{figure}[]
  \begin{center}
    \subfigure[\(\hat{\zeta}_{m}=1\)]{\label{fig:edge-a}\includegraphics[scale=1, width=6.5cm, height=3.5cm]{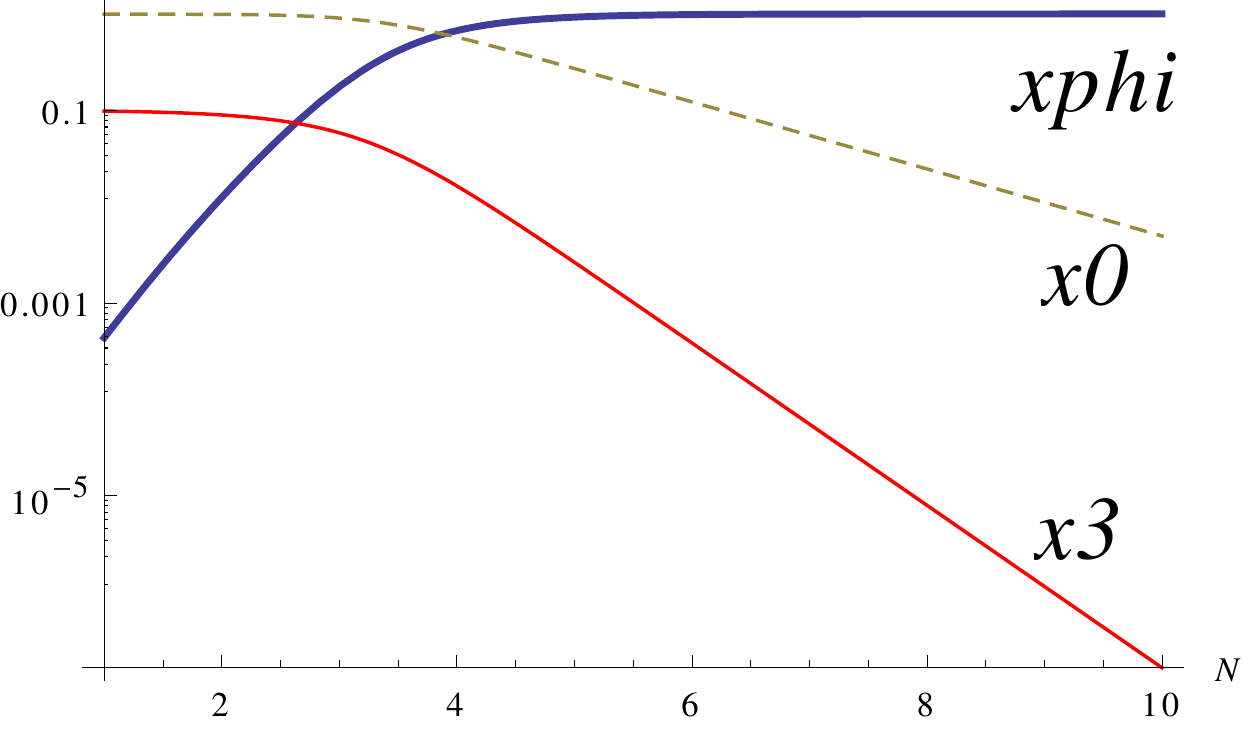}}
    \subfigure[\(\hat{\zeta}_{m}=-1\)]{\label{fig:edge-b}\includegraphics[scale=1, width=6.5cm, height=3.5cm]{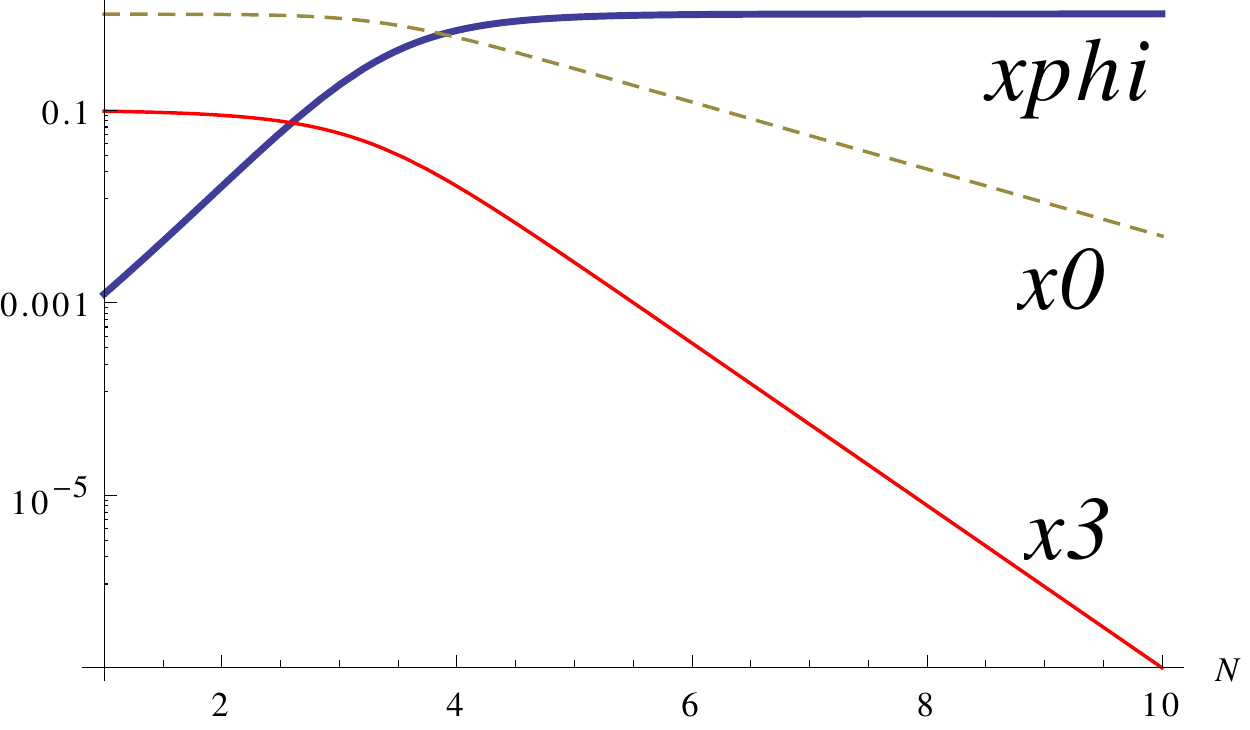}}
  \end{center}
  \caption{Simulations for a universe with dust and a dilatonic ghost condensate scalar field, with \(\omega=6\) and \(\lambda=0.5\) in both cases. The thick blue curve is the magnitude of \(x_{\phi}=x_{1}^{2}(3x_{4}^{2}-1)\), the red of \(x_{3}\) and the dashed yellow of \(x_{0}\). For both simulations we choose initial conditions \(x_{1}(0)=0.01\), \(x_{3}(0)=0.1\) and \(x_{4}(0)=0.8\). In both cases the solution quickly becomes dominated by the kinetic energy of the scalar field and approaches an uncoupled, K-essence attractor solution. The sign of \(\zeta_{m}\) has little effect on the dynamics.}
  \label{fig:edge}
\end{figure}

\begin{figure}[]
  \begin{center}
    \subfigure[\(\hat{\zeta}_{m}=1\)]{\label{fig:edge-a}\includegraphics[scale=1, width=6.5cm, height=3.5cm]{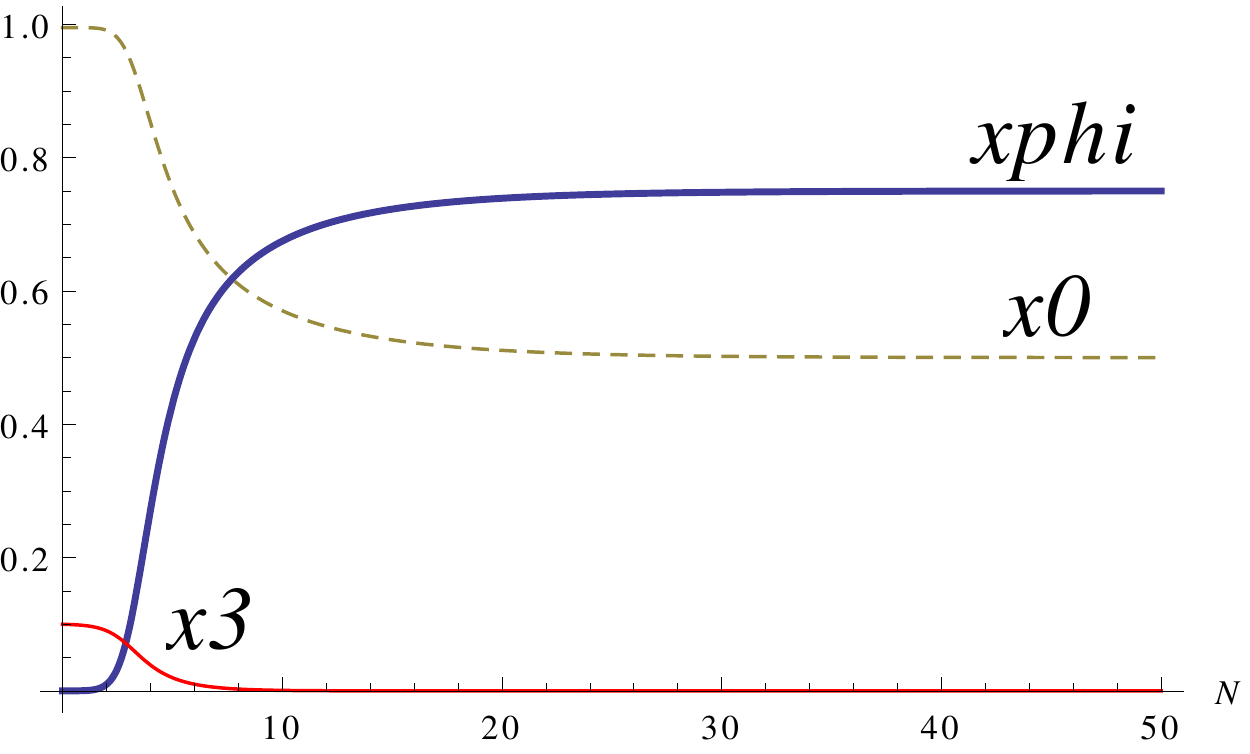}}
    \subfigure[\(\hat{\zeta}_{m}=-1\)]{\label{fig:edge-b}\includegraphics[scale=1, width=6.5cm, height=3.5cm]{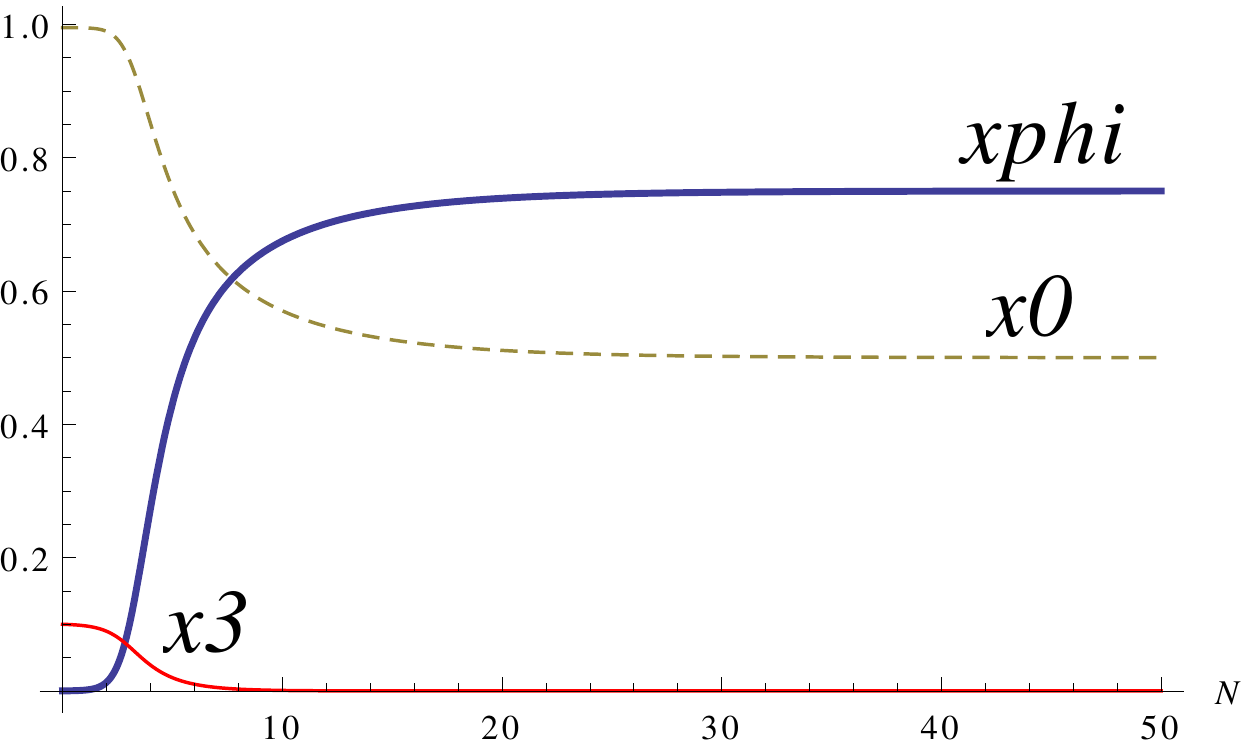}}
  \end{center}
  \caption{Simulations for a universe with dust and a dilatonic ghost condensate scalar field, with \(\omega=6\) and \(\lambda=2\) in both cases. The thick blue curve is the magnitude of \(x_{\phi}=x_{1}^{2}(3x_{4}^{2}-1)\), the red of \(x_{3}\) and the dashed yellow of \(x_{0}\). For both simulations we choose initial conditions \(x_{1}(0)=0.01\), \(x_{3}(0)=0.1\) and \(x_{4}(0)=0.8\). In both cases the solution approaches the scaling solution. The sign of \(\zeta_{m}\) has little effect on the dynamics.}
  \label{fig:edge}
\end{figure}

\section{General models}
In the previous two sections  we have studied the late-time behaviour of our model for two particular choices of \(P(\phi,X)\) that are amenable to dynamical systems analysis. For a general choice of \(P(\phi,X)\) this is not so simple. Despite this, we can give a general argument that any model of this form which hopes to cause acceleration at late times must lead to \(\alpha\) variations which are too large.

For simplicity, and because we are in any event primarily interested in models where the kinetic energy of the scalar field drives acceleration, let us assume the Lagrangian is purely kinetic, so that \(P=P(X)\). In this case the scalar field equation of motion, Eq. \eqref{3.2}, can be written in a simple form:
\begin{equation} \label{6.1}
\frac{d}{dt}(a^{3}\dot{\phi}P_{,X})=-\frac{2}{\sqrt{\omega}}\rho_{m}a^{3}\zeta_{m}e^{\frac{-2\phi}{\sqrt{\omega}}}. 
\end{equation}
Now everything on the right hand side of this equation is a constant except the exponential factor, and this is in general a decreasing function of time. We may therefore expect the right hand side to tend to zero as \(t\rightarrow\infty\), which should imply that \(\dot{\phi}P_{,X}\rightarrow0\) as \(t\rightarrow\infty\). More precisely, we may rewrite Eq. \eqref{6.1} as
\begin{equation} \label{6.2}
\dot{\phi}P_{,X}\propto\frac{1}{a(t)^{3}}\int{}e^{-2\phi(t')/\sqrt{\omega}}dt'. 
 \end{equation}
Now if the late-time attractor is an accelerating scalar-dominated state, or a scaling solution, then clearly the scale factor may expand no slower than the scale factor for the Einstein-de Sitter universe, so \(a(t)^{3}\geq{}t^{2}\) as \(t\rightarrow\infty\). Since \(\phi(t)\) is an increasing function of time\footnote{Note that although \(\phi(t)\) is an increasing function of time it is not necessarily the case that \(\int{}e^{-2\phi(t)/\sqrt{\omega}}dt\) is a decreasing function of time. For instance, if \(\phi\) is asymptotically of the form \(\phi(t)\rightarrow{}n\ln{}t\) then \(\int{}e^{-2\phi(t)/\sqrt{\omega}}dt\rightarrow{}t^{1-\frac{2n}{\sqrt{\omega}}}\), unless \(n=\sqrt{\omega}/2\) in which case it grows logarithmically; when \(n\leq\sqrt{\omega}/2\) this does not tend to zero as \(t\rightarrow\infty\). For most reasonable choices of \(\phi(t)\) it cannot grow faster than \(t^{2}\) though, which suffices to establish the vanishing of the right hand side of Eq. \eqref{6.2} in the limit \(t\rightarrow\infty\).} we thereby expect that the right hand side tends to zero as \(t\rightarrow\infty\). Hence, in an ever expanding universe Eq. \eqref{6.2} implies that \(\dot{\phi}P_{,X}\rightarrow0\), which implies that either \({\phi}\rightarrow\mbox{constant}\) or \(P_{,X}\rightarrow0\) as \(t\rightarrow\infty\). The first choice implies that the kinetic energy of the scalar field vanishes at late time, but if this is so the universe will decelerate. If we wish the scalar field to behave as a K-essence field this possibility must be rejected, so we have \(P_{,X}\rightarrow0\) as \(t\rightarrow\infty\). However, for a non-canonical scalar field the solution of this will in general by \(X\neq0\) unless \(P_{,X}\) contains no constant term.  Indeed, if the solution of this equation was that \(X\rightarrow0\) then the scalar fields kinetic energy would tend to zero, so presumably would be unable to drive acceleration. We conclude then that generically we cannot have that \(\phi\rightarrow\mbox{constant}\) if the scalar field is indeed to perform as a K-essence scalar field.

\section{Conclusions}
In this paper we have studied a new covariant theory of varying-\(\alpha\) in which the fine structure constant evolves through space and time due to a direct coupling between the electromagnetic sector and a non-canonical scalar field. It is a fairly natural development of the BSBM model and its extensions.

In particular, we studied in detail the case where the scalar field is of ghost condensate type, since models of this form are known to lead to cosmic acceleration driven by the kinetic energy of the scalar field. We showed that the system always evolves either to one dominated by the scalar field or a scaling solution. In either case, the scalar field grows at least as fast as a logarithm, which implies that \(\alpha\) grows as a power-law of time. This means models of this form will be generically unacceptable as combined models of varying-\(\alpha\) and K-essence for the same reason BSBM with a quintessence potential is: the variation of \(\alpha\) at late times is too fast to be compatible with terrestrial constraints without fine-tuning \cite{barrow08}. We also argued this is a generic feature of all models which attempt to simultaneously accelerate the universe via a K-essence scalar field and produce \(\alpha\) variations via the same scalar field. Another way of stating this is that if the present observed cosmic acceleration is indeed driven by a K-essence scalar field it cannot couple to the electromagnetic sector.

One way this could be avoided is by introducing two scalar fields, \(\phi\) and \(\Psi\), one of which (\(\phi\)) is massless and couples to the electromagnetic sector and the other (\(\Psi\)) is an uncoupled K-essence scalar field. Since we know that the back-reaction of the \(\phi\) field on the background expansion is negligible the background expansion will be given entirely by the \(\Psi\)-field dynamics, with the growth of \(\alpha\) determined by \(\Psi\) implicitly (since this will control the scale factor). An interesting question would be is this behaviour stable if a direct coupling between the two fields was introduced, or would it destroy the desired behaviour of \(\phi\)?

One of the deficiencies of this study is that we were forced to be fairly restrictive in our choice of \(P(\phi,X)\). This is largely because, unlike the case of a canonical scalar field with a self-interaction potential, it is difficult to cast Eqs. \eqref{3.2}-\eqref{3.4} into a useful autonomous form for a general \(P(\phi,X)\). There are other strategies one might employ for casting the equations into autonomous form, like the strategy in Ref. \cite{wands13}, which may prove useful for different choices of \(P(\phi,X)\).

Even within the class of models defined by the action in Eq. \eqref{3.5} we have not explored fully the range of possibilities, since we set the potential to zero. Allowing for a non-constant potential creates a range of new and interesting behaviour such as bouncing cosmologies \cite{lin11}. We have also not investigated the behaviour of perturbations in these cosmologies. If these models were taken seriously this would need to be addressed separately.

\section*{Acknowledgements}
I thank John Barrow for his advice and guidance on the project, and for his comments on a draft of this paper. I am supported by the STFC.

\end{document}